\documentclass[aps,showpacs,preprintnumbers,final]{revtex4}
\usepackage{amsfonts}
\usepackage{bm}
\usepackage[dvips]{graphicx,color}
\DeclareGraphicsExtensions{.jpg}

\begin{document}

\title{Correpondence between the one-loop three-point vertex \\and the Y- and $\Delta$- electric resistor networks}
\author{A.T.Suzuki\footnote{On Sabbatical leave at Department of Physics, North Carolina State University, Raleigh, NC}}
\affiliation{Instituto de F\'{\i}sica Te\'orica-Universidade Estadual Paulista,\\
Rua Dr. Bento Teobaldo Ferraz, 271 -- Bloco II -- 01140-070 S\~ao Paulo, SP -- Brazil}
\author{}
\affiliation{}
\author{}
\affiliation{}
\date{\today }

\begin{abstract}
Different mathematical methods have been applied to obtain the analytic result for the massless triangle Feynman diagram yielding a sum of four linearly independent hypergeometric functions of two variables $F_4$. These are defined for especific regions of convergence for the ratios of the squares of momentum variables. 
In this paper I work out the diagram and show that that result, though mathematically sound, is not physically acceptable when it is embedded in higher loops - meaning further momentum integrations -  because it misses a fundamental physical constraint imposed by the conservation of momentum, which should reduce by one the total number of linearly independent (l.i.) functions $F_4$ in the overall solution. Taking into account that the momenta flowing along the three legs of the diagram are constrained by momentum conservation, the number of overall l.i. functions that enter the most general solution must reduce accordingly.

To determine the exact structure and content of the analytic solution for the three-point function that can be embedded in higher loops, I use the analogy that exists between Feynman diagrams and electric circuit networks, in which the electric current flowing in the network plays the role of the momentum flowing in the lines of a Feynman diagram. This analogy is employed to define exactly which three out of the four hypergeometric functions are relevant to the analytic solution for the Feynman diagram. The analogy is built based on the equivalence between electric resistance circuit networks of type ``Y'' and ``Delta'' in which flows a conserved current. The equivalence is established via the theorem of minimum energy dissipation within circuits having these structures. 
\end{abstract}

\pacs{11.10.Gh,11.15.Bt,11.55.Bq, 12.38.Bx, 14.70.Dj}
\maketitle







\section{Introduction}

In working out the higher order loops of Feyman diagrams, one has to perform a growing number of momentum integrals. It may happen that the total integration may be performed via sectioning it so that first ``simpler'' lower-loop integrals are performed and then the result embedded in a higher order diagram integral. This method of doing things is not new and have been applied ever since higher loop calculations have been performed in quantum field theory calculations. There are several types of one-loop integrals that occur in these more complex integrands of higher loop diagrams; among which we cite the one-loop self-energy type integrals and one-loop triangle diagram type integrals. In this work I focus my attention in the latter type of integrals.   

\begin{figure}[h]
\centering
\includegraphics[scale=0.5]{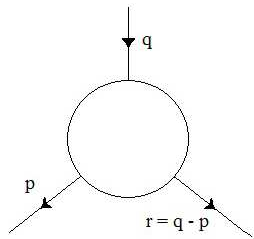}
\caption{One-loop triangle Feynman diagram.}
\label{Figure}
\end{figure}
Massless triangle Feynman diagrams (see Fig.\ref{Figure}) occur in field theoretical perturbative calculations, being one of the primary divergent (one-particle irreducible) graphs for the non-Abelian gauge theories such as Yang-Mills fields. This makes this kind of perturbative calculation the more interesting and necessary once we want to check the higher order quantum corrections for many physical processes of interest. 

Two decades ago, Boos and Davydychev \cite {BoosDavy} were the first to obtain the analytic formulae for the one-loop massless vertex diagram using the Mellin-Barnes complex contour integral representation for the propagators. Later on, Suzuki, Santos and Schmidt \cite {SSS} reproduced the same result making use of the negative dimensional integration method (NDIM) technique of Halliday and Ricotta \cite{HR} for Feynman integrals, where propagators are initially taken to be finite polynomials in the integrand, and once the result is obtained, analytically continued to the realm of positive dimensions.   

These two widely different methods of calculation yielding the same answer lends to the analytic result obtained a degree of certainty as far as the mathematical soundness is concerned. 
The dimensionally regularized ($D=2\omega$) massless triangle integral reads
\begin{equation}
\label{triangle}
I_{\Delta}\left(p^2,\;q^2,\;r^2\right) = \int \frac{d^{D}k}{(k^2+i\epsilon) [(k-p)^2+i\epsilon] [(k-q)^2+i\epsilon]}
\end{equation}
in which $p$ and $q$ are two of the independent external momenta, and $r = q-p$ the momentum for the diagram's third leg. It yields \cite {BoosDavy, SSS}
\begin{eqnarray}
I_\Delta & = & \pi^\omega \left(p^2\right )^{\omega-3} \left \{{\bf A}\,F_4(\alpha, \, \beta, \, \gamma, \, \gamma\hspace{.05cm}';\, x, \, y) \right. \label{A} \\
&&\hspace{1.70cm}+ {\bf B}\,x^{1-\gamma}\,F_4(\alpha+1-\gamma, \, \beta+1-\gamma, \, 2-\gamma, \, \gamma\hspace{.05cm}'; \, x, \, y)\label{B} \\
&&\hspace{1.70cm}+  {\bf C}\,y^{1-\gamma\hspace{.05cm}'}\,F_4(\alpha+1-\gamma\hspace{.05cm}', \, \beta+1-\gamma\hspace{.05cm}', \, \gamma, \, 2-\gamma\hspace{.05cm}'; \, x, \, y)\label{C} \\
&&\hspace{1.70cm}+   {\bf D}\,x^{1-\gamma}y^{1-\gamma\hspace{.05cm}'}\,\left.F_4(\alpha+2-\gamma-\gamma\hspace{.05cm}', \, \beta+2-\gamma-\gamma\hspace{.05cm}', \, 2-\gamma, \, 2-\gamma\hspace{.05cm}'; \,x, \, y) \right\} \label{D1}
\end{eqnarray}
where we have introduced the following definitions for the coefficients:
\begin{eqnarray}
{\bf A} & \equiv & \frac{\Gamma^2(\omega-2)\Gamma(3-\omega)}{\Gamma(2\omega-3)}
\nonumber \\
{\bf B}= {\bf C} & \equiv & \frac{\Gamma(2-\omega)\Gamma(\omega-2)\Gamma(\omega-1)}{\Gamma(2\omega-3)} \nonumber \\
{\bf D} & \equiv & \Gamma^2(2-\omega)\Gamma(\omega-1),
\end{eqnarray}
and the various hypergeometric function parameters and variables are given by:
\begin{eqnarray}
\alpha & = & 1,\nonumber \\
\beta = \gamma = \gamma\hspace{.05cm}' & = & 3 - \omega,\nonumber \\
x & = & \frac{r^2}{p^2}, \nonumber \\
y & = & \frac {q^2}{p^2}. 
\end{eqnarray}

As it stands by itself, the analytic result (\ref{A})--(\ref{D1}) is perfectly sound and good. However, when we consider it as a part of a higher order diagram calculation, it is no longer valid since the variables in the $F_4$'s are constrained by convergence rules which further momentum integration violates. So we need to look for the solution that will allow such further integration without violating convergence constraints. 

To begin our search let us consider the structure of the solution as it stands with four $F_4$'s. From the study of higher transcendental functions, consider the following system of coupled, simultaneous partial differential equations \cite{Appell}:
\begin{eqnarray}
x(1-x)\frac{\partial^2 Z}{\partial x^2}-y^2\frac{\partial^2 Z}{\partial y^2}-2xy\frac{\partial^2 Z}{\partial x \partial y}
+ [\gamma-(\alpha+\beta+1)x]\frac{\partial Z}{\partial x} - (\alpha+\beta+1)y\frac{\partial Z}{\partial y}-\alpha \beta Z =  0 \nonumber \\
y(1-y)\frac{\partial^2 Z}{\partial y^2}-x^2\frac{\partial^2 Z}{\partial x^2}-2xy\frac{\partial^2 Z}{\partial x \partial y}
+ [\gamma\hspace{.05cm}'-(\alpha+\beta+1)y]\frac{\partial Z}{\partial y} - (\alpha+\beta+1)x\frac{\partial Z}{\partial x}-\alpha \beta Z  = 0,
\end{eqnarray}
with variables $x$ and $y$ independent. The most general solution for such a system of differential equations consists of a linear combination of four hypergeometric functions of type $F_4$, as follows \cite{Appell}:
\begin{eqnarray}
Z & = & {\bf A}\,F_4(\alpha, \, \beta, \, \gamma, \, \gamma\hspace{.05cm}';\, x, \, y)\label{A1} \\
&+& {\bf B}\,x^{1-\gamma}\,F_4(\alpha+1-\gamma, \, \beta+1-\gamma, \, 2-\gamma, \, \gamma\hspace{.05cm}'; \, x, \, y)\label{B1} \\
&+& {\bf C}\,y^{1-\gamma\hspace{.05cm}'}\,F_4(\alpha+1-\gamma\hspace{.05cm}', \, \beta+1-\gamma\hspace{.05cm}', \, \gamma, \, 2-\gamma\hspace{.05cm}'; \, x, \, y)\label{C1} \\
&+& {\bf D}\,x^{1-\gamma}y^{1-\gamma\hspace{.05cm}'}\,F_4(\alpha+2-\gamma-\gamma\hspace{.05cm}', \, \beta+2-\gamma-\gamma\hspace{.05cm}', \, 2-\gamma, \, 2-\gamma\hspace{.05cm}'; \,x, \, y). \label{D2}
\end{eqnarray}

Naive comparison between (\ref{A})--(\ref{D1}) and (\ref{A1})--(\ref{D2}) would tempt us to write $Z = I_\Delta$ without much a do. However, we have to remember that for the case $I_\Delta$, variables $x$ and $y$ are not independent ones, since they are connected by momentum conservation: $r=q-p$. Therefore, the most general solution for the underlying system of coupled equations that are associated to the triangle diagram should not have {\it four} independent functions $F_4$ but only {\it three}, on account of the one constraint due to momentum conservation connecting variables $x$ and $y$. 

From a pure mathematical point of view then, the solution for the triangle Feynman diagram must obey this requisite, since not all the four hypergeometric functions in it are linearly independent to each other. 

Therefore, on a sound mathematical basis, when we are interested in processes requiring momentum integration, this constraint must be considered and the result for the triangle diagram integral should be expressed as a combination of only {\it three} linearly independent hypergeometric functions $F_4$. In other words, the result for triangle diagrams that is to be embedded in higher order diagrams must allow for all the momentum integration range, which is forbidden in the four term $F_4$'s format, due to the variables convergence constraint. 

Neither the Mellin-Barnes formulation of Boos and Davydychev nor the NDIM technique employed by Suzuki {\sl et al} to evaluate the massless triangle graph take into account the basic physical constraint imposed on such a diagram due to the momentum conservation flowing in the three legs. Momentum (that is, energy and three-momentum) conservation is one of the basic, fundamental tenets of modern natural sciences, and plays a key role in determining the final form for the analytic result for the massles one-loop vertex correction without convergence constraint forbiding further momentum integration.

\begin{figure}[h]
\centering
\includegraphics[scale=0.5]{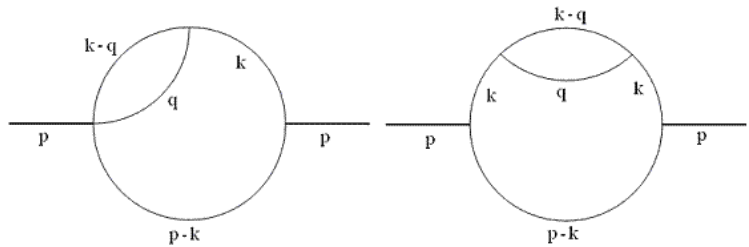}
\caption{Two-point two-loop Feynman diagrams: Fying-saucer side-view and front-view.}
\label{Figure2}
\end{figure}

One of the simplest examples where the four-term hypergeometric combination (\ref{A}) --(\ref{D1}) above mentioned does not reproduce the correct result is when it is embedded in a higher two-loop order calculation \cite{SSB}. Consider, for example, the diagrams of Fig.\ref{Figure2}. 

These two two-loop Feynman diagrams have similar loop momentum integrand structures, namely, 
\begin{equation}
\label{fs}
I_{FS}^{(2)} = \int \int \frac{d^Dk d^Dq}{(k^2)^{a}q^2[(k-p)^2][(q-k)^2]}
\end{equation}
in which we omitted the infinitesimals ($i\epsilon)$ associated to each propagator in the denominator and $a$ can be either $1$ or $2$ depending on which diagram one considers.  

We can perform the double integration in two ways. The easiest one is doing the $q$ integration first and then the $k$ integration, 
\begin{equation}
\label{se}
I_{FS}^{(2)} = \int  \frac{d^Dk}{(k^2)^{a}[(k-p)^2]}\int \frac{d^D q}{q^2[(q-k)^2]}
\end{equation}
so that both integrands are of the one-loop self-energy type, and the results are:
\begin{eqnarray}
\label{selfenergy}
I_{FS}^{(2)} = \pi^{2\omega} 
\left \{
\begin{array}{l}(p^2)^{2\omega - 4} \displaystyle\frac{\Gamma^3(\omega-1)\Gamma(2-\omega)\Gamma(2\omega-3)\Gamma(4-2\omega)}{\Gamma(2\omega-2)\Gamma(3-\omega)\Gamma(3\omega-4)} ,\qquad {\rm for}\;\;a=1,  \\ 
\\
(p^2)^{2\omega-5}\displaystyle\frac{\Gamma^3(\omega-1)\Gamma(2-\omega)\Gamma(2\omega-4)\Gamma(5-2\omega)}{\Gamma(2\omega-2)\Gamma(4-\omega)\Gamma(3\omega-5)}, \qquad {\rm for}\;\;a=2.
   \end{array}\right.
\end{eqnarray}

The result (\ref{selfenergy}) should also be obtainable from doing $k$ integration first and then $q$ integration at last:
\begin{equation}
I_{FS}^{(2)} = \int \frac{d^Dq}{q^2}\int \frac{d^Dk}{(k^2)^a[(k-p)^2][(q-k)^2]}
\end{equation}

It is clear that this path is harder to travel, since the first integrand is of the one-loop triangle diagram type, (\ref{triangle}). 

It is here that the plot thickens, because those results in (\ref{selfenergy}) can never be achieved by using the triangle diagram as expressed in terms of (\ref{A}) - (\ref{D1}). For neither of the diagrams the four-term solution for the triangle when embedded into the two-loop graph yields the correct answer. The right result can only be achieved when we use a three-term triangle solution embedded into the two-loop structure \cite{SSB}. What is the problem? I argue that the problem stems from the fact that these four terms  (\ref{A}) - (\ref{D1}) are not correct because they are not linearly independent solutions; physical momentum-energy conservation law must be considered, and this must reduce by one the overall number of $F_4$'s that should enter into the most general solution for $I_\Delta$ in equation (\ref{triangle}).

How do we go about in reducing those four terms into three? We could use the analytic continuation formula for the $F_4$ function \cite{Appell}:
\begin{eqnarray}
\label{anacon}
F_4(a,b,c,c';z,w)& = & \frac{\Gamma(d)\Gamma(b-a)}{\Gamma(b)\Gamma(d-a)}(-w)^{-a} F_4(a,a+1-d,c,a+1-b;\frac{z}{w},\frac{1}{w})\nonumber \\
& + & \frac{\Gamma(d)\Gamma(a-b)}{\Gamma(a)\Gamma(d-b)}(-w)^{-b}F_4(b+1-d,b,c,b+1-a;\frac{z}{w},\frac{1}{w})
\end{eqnarray}
in order to reduce the number of independent functions from four to three. 

However, for the overall four hypergeometric functions $F_4$ that appear in $I_\Delta$, one has at least two possibilities to pair them two by two according to (\ref{anacon}): Either combining (\ref{A}) and (\ref{C}) or combining (\ref{B}) and (\ref{D1}).

Making a suitable change of variables and redefining conveniently the various parameters of the hypergeometric function, we have, for example, pairing (\ref{B}) and (\ref{D1}) according to (\ref{anacon}):
\begin{eqnarray}
\label{anacont}
\left. x^{1-\gamma}\right\{&\!\!\!\!\!\!\!{\bf B}&\!\!\!\!\!\!\!F_4(\alpha+1-\gamma,\beta+1-\gamma,2-\gamma,\gamma\,';x,y)
+\left. {\bf D}y^{1-\gamma\,'}F_4(\alpha+2-\gamma-\gamma\,',\beta+2-\gamma-\gamma\,',2-\gamma,2-\gamma\,';x,y)\right \}\nonumber\\
&=&\Omega({\bf B},{\bf D})F_4\left(\beta+2-\gamma-\gamma\,',1+\beta-\gamma,2-\gamma,1-\alpha+\beta;\frac{x}{y},\frac{1}{y}\right)
\end{eqnarray}
where $\Omega({\bf B},{\bf D})$ is a new coefficient which depends on the previous ones ${\bf B}$ and ${\bf D}$. 

Another possibility would be to make a similar suitable change of variables and redefinition of parameters in a convenient way so as to pair (\ref{A}) and (\ref{C}) according to (\ref{anacon}). Either way we accomplish the reduction to three linearly independent $F_4$'s without convergence constraint that forbids further momentum integration, but which one must we choose?

In other words, which three of those four $F_4$'s contained in the result $I_\Delta$ should be considered cannot be determined by momentum conservation only; it needs another physical input to this end.  

In order to determine the exact structure and content of the analytic result for the triangle diagram I employ the analogy that exists between Feynman diagrams and electric circuit networks. The conclusion of the matter can be summarized as follows: since there is a constraint (call it either a initial condition or a boundary condition) to the momenta flowing in the legs of a triangle diagram, it means that the overall analytic answer obtained via Mellin-Barnes or NDIM technique, which comes as a sum of four linearly independent hypergeometric functions of type $F_4$, in fact must contain only three linearly independent ones. Which three of these should be is determined by the momentum conservation flowing through the legs of the diagram, {\sl and}, by the circuit analogy, by the equivalence between the ``Y-type'' and ``$\Delta$-type'' resistance networks through which {\sl conserved} electric current flows.          
  
\section{Electric circuitry and Feynman diagrams}

Although known for a long time and mentioned sometimes in the literature of field theory, the analogy between electric circuit networks and Feynman diagrams, more often than not stays as a mere curiosity or at a diagrammatic level in which the drawing is only a representation that helps us out in the ``visualization'' of a given physical process. 

There is, however, a few exceptions to this. The earliest one that I know of is by Mathews \cite {Mathews} as early as 1958 where he deals with singularities of Green's functions and another one by Wu \cite {Wu} in 1961, where he gave an algebraic proof for several properties of normal thresholds (singularities of the scattering amplitudes) in perturbation theory. 

What I am going to do here is just to state the well-known result for the equivalence between ``$Y$''- and ``$\Delta$-circuits'' which is used, for example, in analysing the ``Kelvin bridge'' in relation to the more commom ``Wheatstone brigde''. The equivalence I consider is for the network of resistors, where energy is dissipated via Joule effect. Then with this in hands I argue which three out of the overall four hypergeometric functions $F_4$ present in the one-loop vertex obtained via Mellin-Barnes or via NDIM should be considered as physically meaningful.  

Note that the equivalence between those circuits of resistors can be established based on the conservation of energy flowing in their internal lines. 

\begin{figure}[!h]
\centering
\includegraphics[scale=0.5]{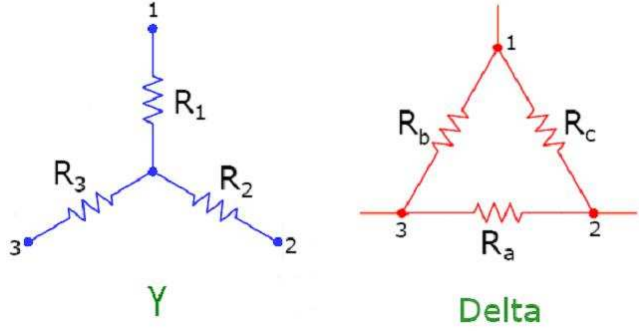}
\caption{``Y-type'' and ``Delta-type'' resistor networks.}
\label{Figure3}
\end{figure}

Consider then the ``$Y$''- and ``$\Delta$-circuits'' of Fig.\ref{Figure3}.

These two network of resistors are related to each other; for example, the ``Delta-circuit'' can be made equivalent to the ``Y-circuit'' when resistors obey the following equations \cite {Farago}:
\begin{eqnarray}
R_a&=&R_2+R_3+\frac{R_2 R_3}{R_1} \nonumber\\
R_b&=&R_1+R_3+\frac{R_1 R_3}{R_2} \label{Ra}\\
R_c&=&R_1+R_2+\frac{R_1 R_2}{R_3} \nonumber
\end{eqnarray}

In electric circuits a general theorem guarantees that the current density in a conductor distributes itself in such a way that the generation of heat is a minimum \cite{Smythe}. Strictly speaking, since I consider the ``Y''- and ``Delta-networks'' by themselves and not embedded within an electrical circuit, I cannot consider the minimum of power generation, but still I can consider the equivalence between the power generated in one of them --- say ``Y-circuit'' --- as compared to the power generated in the other one, the ``Delta-circuit''.     

Labelling the currents that flow through resistors $R_1$, $R_2$ and $R_3$ of the ``Y-circuit'' by $p$, $q$ and $r$, respectively, by reason of current conservation we have that, for example \footnote[1]{Of course, we can take either two of those as independent ones, such that the dependent current can be expressed also as either $q = p+r$ or $p = q-r$}
\begin{equation}
r = q - p
\end{equation}

Each of these currents will generate heat according to Joule's law and the corresponding power will be given by Ohm's law. So in each leg, the product of the resistance times the square of the current will give the power generated by the current flowing through it. The overall power generated in the ``Y-network'' is therefore: 
\begin{equation}
\label{Ypower}
p^2R_1+q^2R_2+r^2R_3=P_{\rm total}
\end{equation}
Dividing (\ref{Ypower}) respectively by $p^2$, $q^2$ and $r^2$ we have the following system of equations:
\begin{eqnarray}
R_1+y R_2+x R_3&=&\frac{1}{p^2}P_{\rm total}\equiv {\bf R}_{\rm p} \label{D}\\
\frac{1}{y}R_1+R_2+z R_3&=&\frac{1}{q^2}P_{\rm total}=\frac{1}{y}{\bf R}_{\rm p}\label{E}\\
\frac{1}{x}R_1+\frac{1}{z}R_2+R_3&=&\frac{1}{r^2}P_{\rm total}=\frac{1}{x}{\bf R}_{\rm p} \label{F}
\end{eqnarray}
where for convenience I have defined  $z\equiv \displaystyle\frac{x}{y} \equiv \displaystyle\frac{r^2}{q^2}$ and ${\bf R}_{\rm p} \equiv \displaystyle\frac{1}{p^2}P_{\rm total}$. 

Multiplying the last equation (\ref{F}) by $x$ and comparing with (\ref{D}),
\begin{eqnarray}
R_1+y R_2+x R_3&=&\frac{1}{p^2}P_{\rm total}\equiv {\bf R}_{\rm p} \label{G}\\
R_1+\frac{x}{z}R_2+x R_3&=&\frac{x}{r^2} P_{\rm total}={\bf R}_{\rm p} \label{H}
\end{eqnarray}
it follows that the coefficient of $R_2$ is $y = \displaystyle\frac{x}{z}$.

In a similar manner, multiplying (\ref{D}) by $\displaystyle\frac{1}{y}$ and comparing with (\ref{E}),
\begin{eqnarray}
\frac{1}{y}R_1+R_2+\frac{x}{y} R_3&=&\frac{1}{y p^{2}}P_{\rm total}\equiv \frac{1}{y}{\bf R}_{\rm p} \label{I}\\
\frac{1}{y}R_1+R_2+z R_3&=&\frac{1}{q^{2}}P_{\rm total}=\frac{1}{y}{\bf R}_{\rm p}\label{J}
\end{eqnarray}
it follows that the coefficient of $R_3$ is $z=\displaystyle\frac{x}{y}$.

Finally, multiplying (\ref{F}) by $yz$ and comparing with (\ref{D}),
\begin{eqnarray}
R_1+y R_2+x R_3&=&\frac{1}{p^{2}}P_{\rm total}\equiv {\bf R}_{\rm p} \label{K}\\
\frac{yz}{x}R_1+y R_2+yz R_3&=&\frac{yz}{ r^{2}}P_{\rm total}=\frac{yz}{x}{\bf R}_{\rm p} \label{L}
\end{eqnarray}
it follows that the coefficient of $R_1$ is $1=\displaystyle\frac{yz}{x}$, showing the thorough consistency among these results. 

On the other hand, in the ``Delta-network'', let $I_a$ label the current flowing through resistance $R_a$, so that the power generated in this resistor is, using (\ref{Ra})
\begin{equation}
P_{\rm a} = I_{a}^2\,R_a = I_{a}^2\,\left \{\frac{R_2\,R_3}{R_1}+R_2+R_3\right \}
\end{equation}

Now, as the resistances follow the equivalence defined by (\ref{Ra}), the current $I_a$ that flows through $R_a$ must be such that:
\begin{equation}
I_a^2R_a \equiv p_a^2\,R_{1,a}+q_a^2\,R_{2,a}+r_a^2\,R_{3,a}
\end{equation}
where
\begin{eqnarray}
R_{1,a}&=&\frac{R_2\,R_3}{R_1}\nonumber\\
R_{2,a}&=&R_2\\
R_{3,a}&=&R_3 \nonumber
\end{eqnarray}
so that the corresponding currents flowing through them must be ``scaled'' accordingly as
\begin{eqnarray}
p_a^2 &=&\frac{yz}{x}\,p^2=p^2 \nonumber\\
q_a^2&=&yq^2\\
r_a^2&=&zr^2
\end{eqnarray}

Therefore, the solution for the one-loop triangle Feynman diagram must be such that it contains three terms which are proportional to ``currents'' $p^2$, $yq^2$ and $zr^2$. This solution is exactly achieved when I combine (\ref{B}) and (\ref{D1}) in a similar way as was done in (\ref{anacont}). Combination of (\ref{A}) and (\ref{C}) as in (\ref{anacont}) leaves the term (\ref{D1}) in the solution, which is proportional to $xyp^{2}$, and therefore not suitable since it violates the equivalence above demonstrated.  

Explicit analytic solution for the one-loop massless triangle Feynman diagram reads therefore
\begin{eqnarray}
\label{result}
\Sigma_{\rm one-loop}(p,q,r)&=& \pi^\omega (p^2)^{\omega-3}\,{\bf \Gamma_p}\,F_4(1,3-\omega,3-\omega,3-\omega;x,y)\nonumber\\
&+&\pi^\omega y(q^2)^{\omega-3}\,{\bf \Gamma_q}\,F_4(1,\omega-1,3-\omega,\,\omega-1;x,y)\\
&+&\pi^\omega z(r^2)^{\omega-3}\,{\bf \Gamma_r}\,F_4\left(1,\,\omega-1,\omega-1,3-\omega;z=\frac{x}{y},\,\frac{1}{y}\right)\nonumber
\end{eqnarray}
where $${\bf \Gamma_p}=-{\bf \Gamma_q}=-{\bf \Gamma_r}=\frac{\Gamma^2(\omega-2)\Gamma(3-\omega)}{\Gamma(2\omega-3)}$$ with $\omega$ being the dimensional regularization parameter. One notes that the variables for the last $F_4$, though related to those that appear in the first and second $F_4$'s, differ from them. 

This construction is particularly important for embedded triangles of higher order Feynman diagrams, since the expression (\ref{result}) bears in it two-variable hypergeometric functions of different types, namely, $x$ with $y$ and $z$ with $y^{-1}$ which covers different regions of convergence. 

\section{Conclusion}

We have seen that the previous result given in the literature for the massless one-loop vertex correction of Feynman diagrams in a general field theory $I_\Delta$ with four linearly independent hypergeometric functions of type $F_4$ can be compared to the most general solution of a system of coupled differential equations for higher transcendental functions. These pressuposes that the variables are independent ones. Since the result $I_\Delta$ has exactly the same structure as the most general solution for the system of differential equations, it follows that the result for the vertex correction can be thought of as having a underlying system of differential equations that it must obey. However, contrary to purely mathematical system of differential equations of two variables, the momentum flowing through the Feynman diagrams obeys a physically important conservation law, which connects the variables $x$, $y$ and $z$ among themselves, via, e.g., $r=q-p$ relation. So, the most general result for it should not have four independent two variable hypergeometric functions $F_4$, but only three.

Using the basic physical principle of momentum conservation I deduced the correct analytic result for the one-loop massless triangle Feynman diagram in terms of three linearly independent hypergeometric functions $F_4$. The reason why there must be only three linearly independent functions in the solution is due to the fact that the variables $x$ and $y$ are such that implicit in them is a momentum conservation constraint that must be taken into account properly. The set of which three linearly independent functions $F_4$ should be is given by another physical input, deduced in analogy to the equivalence between ``Y''- and ``Delta-network'' of electrical resistance circuits. With the conservation of momentum correctly taken into account, the final result for the mentioned Feynman diagram is the physically correct and relevant analytic solution to the problem, having only a combination of three linearly independent $F_4$ functions in it.

Moreover, of particular importance is this three-term result when vertex corrections are inserted in more complex Feynman diagrams. If one takes the four-term result $I_\Delta$ with all the hypergeometric functions with same variables $x$ and $y$, when this is further integrated in loop momentum $k$ of higher order diagrams, it cannot be integrated on the whole domain of integration $-\infty<k<+\infty$ since all the $F_4$'s entering the solution $I_\Delta$ has the same and definite domain of convergence. Not so with the (\ref{result}) where the third $F_4$ allows for the complementary domain region of momentum integration. To understand speciffically which regions of convergence are for each of the two-variable hypergeometric functions $F_4$ see for example \cite{Appell}. Having said that, whenever one has to deal with calculating multiloop Feynman diagrams in which the one-loop vertex correction is a subdiagram thereof, the correct result to embed in the integrand of the multiloop diagram integral {\it must} be the one given in (\ref{result}).   

Finally, for the sake of completeness, I would like to emphasize that two other symmetric results can be obtained just by cyclic permutations among $(p,q,r)$ momentum variables, following the original symmetry of the triangle diagram.

\vspace{1cm}
{\tt Acknowledgments} Work dedicated to Mitiko, my beloved wife, and to our children: Tamie, on her 15th birthday, Tise on her 13th birthday and Timothy on his 7th birthday.

\end{document}